\documentclass[useAMS,usenatbib]{mn2e}
\usepackage{latexsym,graphicx,natbib}
\usepackage{amsmath,subfigure}

\def\fun#1#2{\lower3.6pt\vbox{\baselineskip0pt\lineskip.9pt
  \ialign{$\mathsurround=0pt#1\hfil##\hfil$\crcr#2\crcr\sim\crcr}}}

\def\gap{\mathrel{\mathpalette\fun >}}

\def\msun{{\rm M_{\odot}}}

\def\kms{{\rm\,km\,s^{-1}}} 
\def\beq{\begin{equation}}
\def\eeq{\end{equation}}

\newcommand{\myr}{\ensuremath{\mathrm{Myr}}}
\newcommand{\kpc}{\ensuremath{\mathrm{kpc}}}
\newcommand{\pc}{\ensuremath{\mathrm{pc}}}

\newcommand\NMS{\ensuremath{N_\mathrm{MS}}}
\newcommand\fall{\ensuremath{f_\mathrm{all}}}
\newcommand\rvir{\ensuremath{R_\mathrm{vir}}}

\newcommand\Ul{\ensuremath{\mathrm{U}_\mathrm{l}}}
\newcommand\Um{\ensuremath{\mathrm{U}_\mathrm{m}}}
\newcommand\Ut{\ensuremath{\mathrm{U}_\mathrm{t}}}
\newcommand\fM{\ensuremath{f_\mathrm{M}}}
\newcommand\fIMF{\ensuremath{f_\mathrm{IMF}}}
\newcommand\fCMP{\ensuremath{f_\mathrm{CMP}}}

\newcommand{\revised}[1]{#1}

\title[Reconstructing the Arches]{Reconstructing the Arches I: Constraining the Initial Conditions}

\author[Harfst et al.]{S.~Harfst$^{1,2}$\thanks{harfst@strw.leidenuniv.nl} 
                       and S.~Portegies Zwart$^{1}$ 
                       and A.~Stolte$^{3}$\\
$^1$Leiden Observatory, Leiden University, PO Box 9513, 2300 RA Leiden, The Netherlands\\
$^2$Department of Astronomy \& Astrophysics, Technical University of Berlin, 10623 Berlin, Germany \\
$^3$1. Physikalisches Institut, University Cologne, Z\"ulpicher Str. 77, 50937 K\"oln, Germany\\}

\begin{document}

\maketitle

\begin{abstract}
  We have performed a series of $N$-body simulations to model the
  Arches cluster. Our aim is to find the best fitting model for the
  Arches cluster by comparing our simulations with observational data
  and to constrain the parameters for the initial conditions of the
  cluster. By neglecting the Galactic potential and stellar evolution,
  we are able to efficiently search through a large parameter space to
  determine e.g. the IMF, size, and mass of the cluster. We find, that
  the cluster's observed present-day mass function can be well
  explained with an initial Salpeter IMF. The lower mass-limit of the
  IMF cannot be constrained well from our models. In our best models,
  the initial total mass down to a mass limit of $0.5\,\msun$ is $(4.9
  \pm 0.8)\cdot10^4\,\msun$. The initial virial radius of the cluster
  is $0.77 \pm 0.12\,\pc$. A concentration parameter of the initial
  King model $W_0 = 3$ gives the best results.
\end{abstract}


\section{Introduction}

The Arches cluster is one of only a few young and massive starburst
clusters in the Milky Way. Its location at a projected distance of
less than $30\,\pc$ from the Galactic centre and an age of only $\sim
2.5\,\myr$ \citep{FNG02,NFH04} make this cluster a unique object for
studying star formation and dynamical processes in the centre of
galaxies \citep{PZMG10}.

The observed present-day mass of the Arches cluster within $R =
0.4\,\pc$ has been estimated with $\sim 1-2 \cdot 10^4\,\msun$
\citep{FKM99,ESM09}.  With this mass a cluster will not survive long
in the Galactic centre environment and evaporate on a time scale maybe
as fast as $\sim 10\,\myr$ \citep{KML99,PZMM02}. The initial mass of
the cluster has been determined from $N$-body simulations, however,
different results have been obtained by different authors:
\citet{KFL00} found that their best model for the Arches cluster had a
total mass of about $2\cdot 10^4\,\msun$; \citet{PZMM02}, on the other
hand, came to the conclusion that the cluster was initially more
massive than $\sim4\cdot 10^4\,\msun$.

The initial mass function (IMF), a key aspect of star formation, seems
to be uniform throughout the universe \citep{BCM10}. This universal
IMF can be described by the power-law found by \citet{S55} for stars
in the solar neighbourhood and is valid from $0.5\,\msun$ to the
highest masses. Below $0.5\,\msun$, the IMF is significantly flattened
\citep[e.g.][]{Kroupa02}.

\begin{figure}
\includegraphics[width=84mm,angle=0]{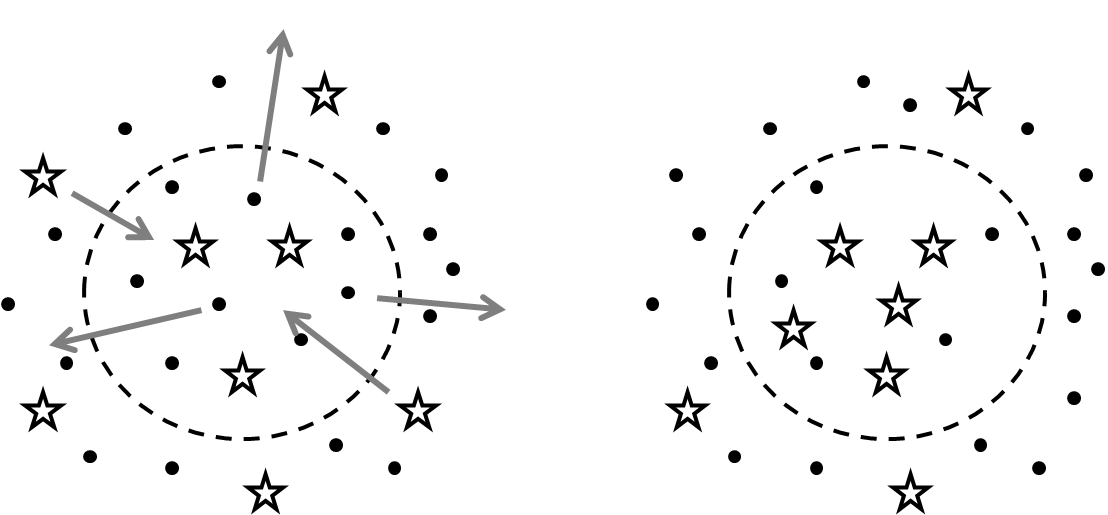}
\caption{Schematic view on the cluster mass function and
  evolution. Low- and high-mass stars are shown by points and star
  symbols, respectively. The dashed circle indicates a radial
  observational limit. In the left image, the cluster is at $t=0$ with
  arrows denoting the effects of dynamical evolution.  The right
  images shows the mass-segregated cluster.}
\label{fig:IMFMS}
\end{figure}

Determining the IMF of young clusters from observations is not a
straight-forward process. Uncertainties can arise from the measured
luminosities, the estimated age of the cluster, the completeness of
the observed sample, and the stellar evolution models. In addition,
the non-linear dynamical evolution of the cluster has to be taken into
account as shown in Fig.~\ref{fig:IMFMS}: as the star cluster evolves,
more massive stars (star symbols) will move towards the cluster centre
and low-mass stars (points) in the opposite direction (indicated by
the arrows in the left image). If the detection of cluster members is
radially limited (dashed circle), it will result in an observed mass
function (MF) in the mass-segregated cluster (right image) that is
different from the IMF. This effect is visualised in
Fig.~\ref{fig:IMFMS} by the ratio of low- to high-mass stars inside
the dashed circles before and after mass segregation.

In case of the Arches cluster, observations have revealed that the
slope of the observed MF for $R \leq 0.4\,\pc$ is significantly
flattened with $\Gamma \approx -0.9 \pm 0.15$ with respect to the
standard Salpeter IMF ($\Gamma = -1.35$) \citep{SBG05,SGB02,FKM99},
and therefore the Arches cluster has been regarded as a possible case
against the universality of the IMF. More recently, however,
\citet{ESM09} derived a slope of $\Gamma = -1.1 \pm 0.2$ in $R <
0.4\,\pc$ and concluded that a standard Salpeter IMF cannot be ruled
out for Arches.  In addition to the radial variation in $A_V$, these
authors also accounted for differential extinction variations, which
can severely affect the incompleteness and may have biased the earlier
results. Large uncertainties in the slope still remain, revealing the
necessity to compare the observed cluster MF with simulations.

In addition to the flattened slope, there has been some debate whether
the IMF of Arches is truncated at the low-mass end as the result of
the extreme conditions at the Galactic centre where the cluster has
formed. Possible evidence for a turn-over in the present-day MF was
reported by \citet{SBG05}, who determined a low- and intermediate-mass
depleted MF in the cluster core ($R < 0.2\,\pc$) with a turn-over at
$6-7\,\msun$. This truncation in the MF was not seen by
\citet{KFK06}. They only found a local bump in the MF at
$\sim6\,\msun$.  Even if the MF is truncated at the low-mass end, it
remains unclear whether this would be the result of a truncated IMF or
a dynamical effect such as tidal stripping of low-mass stars.




With the aim to account for tidal stripping and mass loss in the
Galactic centre potential, several studies have been done to determine
the global IMF of the Arches cluster using numerical simulations,
again coming to different conclusions: the model favoured by
\citet{KFL00} started with flat IMF with a slope of $\Gamma = -0.75$
close to the observed one.  \citet{PZMM02} found that the observed MF
is the result of the dynamical evolution of the cluster and
observational selection effects (namely radius-limited
selection). They argue that the observed flat MF in the cluster core
is therefore consistent with a global Salpeter IMF. The same effect is
seen by \citet{KFK06}, however they suggest the slope of the IMF was
$-1$ to $-1.1$, slightly shallower than Salpeter.

The Arches cluster also exhibits other clear signs of mass
segregation. The slope of the observed MF for stars in different
annuli changes with distance from the cluster centre. Towards the
centre, the slope becomes shallower and further out the slope is
closer to Salpeter \citep{KFK06,SBG05}. Most recently, \citet{ESM09}
have reported $\Gamma = -0.9$ for $R<0.2\pc$ and $\Gamma = -1.3$ in
the $0.2-0.4\pc$ annulus. \citet{PZGC07} have found the same
characteristics in numerical $N$-body models and concluded that the
central flattening is the result of mass segregation. Furthermore,
they claimed that the MF near the centre of the Arches cluster can be
best described by a broken power-law, with a turning point at
$5-6\msun$ (at the position of the bump reported by
\citet{KFK06}). Based on these findings, they determined that the
Arches cluster is about half-way to core collapse.



Despite the wealth of detailed observational data, large uncertainties
regarding some of the properties of the Arches cluster remain, most
importantly in the slope of the observed MF. Numerical simulations
have been used to better understand the observations, but no fiducial
model has emerged from these studies so far. In this paper, we want to
expand previous numerical studies in order to reconstruct the initial
properties of the Arches cluster. For this purpose, we have devised a
new systematic method for comparing the results of $N$-body
simulations with the present-day observations and apply this method to
the Arches cluster. \revised{In addition to the mass function, our
  comparison method also considers the radial mass distribution, which
  has not been done previously.}  Another motivation for revisiting
the Arches cluster is the recent determination of the cluster's proper
motion by \citet{SGM08}. With this knowledge, the orbit of the cluster
can be constraint, and a circular orbit used in some numerical studies
\citep{KFL00,PZMM02} can be ruled out.

In our $N$-body simulations, we model the Arches cluster on a
star-by-star basis and, for the first time, systematically explore the
parameter space to find the best set of initial conditions describing
the Arches cluster. We vary parameters determining the initial mass,
size, and concentration of the cluster. In addition, we also test
which IMF can best explain the MF observed today. The total number of
free parameters in our models is five, and a large number of
simulations is required to search the full parameter space. Therefore,
we decided to neglect the Galactic potential and stellar evolution for
now. The aim of this paper is to introduce our comparison method and
to constrain the initial conditions of the cluster. In a subsequent
paper, we will use these results for simulations that include the here
neglected processes. \revised{In particular, we will be able to extend
  the previous orbit analysis by \citet{SGM08}, who employed a
  leap-frog integration of a point-mass cluster in a logarithmic
  potential. Our new models will provide additional constraints on the
  orbit based on the tidal effects on the cluster, which in turn will
  give us information about the birth place of the Arches cluster and
  its fate.}

\revised{The initial conditions of the Arches cluster are of
  particular interest as the cluster was born in an extreme
  environment and is one of the few local examples of a star
  burst. NGC 3603 is the closest relative, with comparable central
  density and total mass, and with a similar observed top-heavy MF
  \citep{HEM08}. Quintuplet is another young massive cluster in the
  Galactic centre \citep{FMM99} and at an age of $\sim4\,\myr$ could
  be regarded as an older brother to Arches. The close location of
  Arches and Quintuplet suggest a prefered locus for the formation of
  these clusters within the central molecular zone.  Insights on the
  formation locus would help us to understand the build-up of the
  inner bulge and the very massive, dense stellar population in the
  centre of the Galaxy. In addition, if cloud collisions are required
  to explain the dynamics of these clusters, this may limit the number
  of clusters capable of forming in the inner Galaxy, and may also be
  extrapolated to the central molecular rings of external galaxies.}

The paper is outlined as follows: in Sec.~\ref{sec:obsdata}, we
describe the observational data we use. Then we explain the parameters
for the cluster model in Sec.~\ref{sec:clustermodel} and the
simulations and their results in ~\ref{sec:simulations}. We conclude
and summarise in Sec.~\ref{sec:conclusions} and
Sec.~\ref{sec:summary}.

\section{Observational data}
\label{sec:obsdata}

We use data from observations by \citet[][NACO data hereafter]{SBG05}.
The NACO data has been taken using the ESO VLT AO system NAOS and the
CONICA near-infrared camera in two wave bands, $H$ and $K_s$. The
field of view is $\sim25\arcsec$ squared or $1\,\pc^2$ (we assume a
distance of $8\,\kpc$ to the cluster) with a resolution of
$\sim0.\arcsec 08$ or $0.003\,\pc$ at the distance of the Arches.

\begin{figure}
\includegraphics[width=84mm,angle=0]{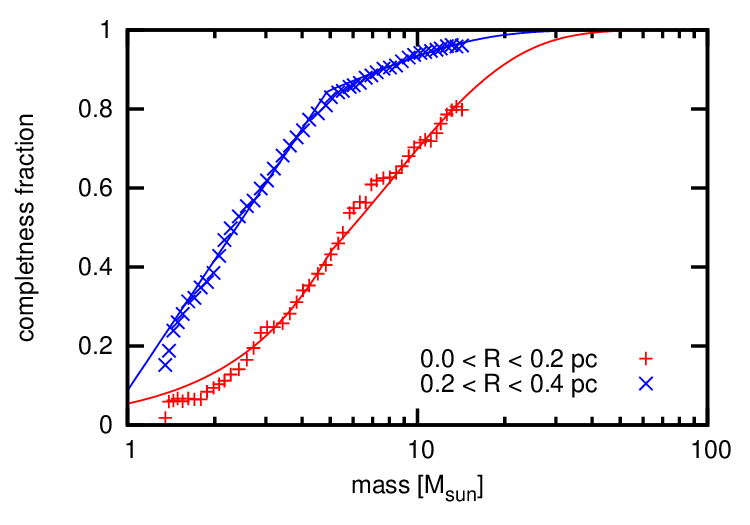}
\caption{The completeness fraction in two radial bins as function of
  initial stellar mass. The data points (blue crosses and red pluses)
  are from the analysis of \citet{SBG05}.  The two full lines show a
  fit to the data. }
\label{fig:inc}
\end{figure}

The total data set consists of $\sim2200$ stars belonging to the
cluster and the field. In order to find the cluster stars, we apply
the same colour selection as \citet{SBG05} (see their Fig.~2),
retaining $\sim1500$ cluster member candidates. We then fit the
$K$-band magnitudes (corrected for the observed radial variation in
extinction \citep{SGB02}) against a $2.5\,\myr$ Geneva main-sequence
isochrone \citep{LS01}, assuming solar metallicity for Arches
\citep{NFH04}. From this we get both present-day and initial stellar
masses for each star in the sample. It should be noted, that the
cluster age used here is different from the age used in \citet{SBG05}
and therefore stellar masses have been re-derived for our preferred
cluster age.


Because the incompleteness of the data due to crowding effects
increases significantly for stars towards the low-mass end we reduced
the sample further to about $300$ stars by selecting only massive
stars with $m>10\,\msun$. With our chosen lower mass limit, the data
is 80\% complete or better for any star mass and position, as a
detailed analysis by \citet{SBG05} has shown. Fig.~\ref{fig:inc} shows
the completeness fraction as a function of initial stellar mass for
two different radial bins. We have fitted the results of \citet[][blue
crosses and red pluses in Fig.~\ref{fig:inc}]{SBG05} and also
extrapolated for stellar masses above $\sim15\,\msun$. In the
following analysis, we use this information to either correct the
observational data or by randomly removing stars from our models. Note
that only a few stars are added or removed by this correction for
$m>10\,\msun$.

\begin{figure}
\includegraphics[width=84mm,angle=0]{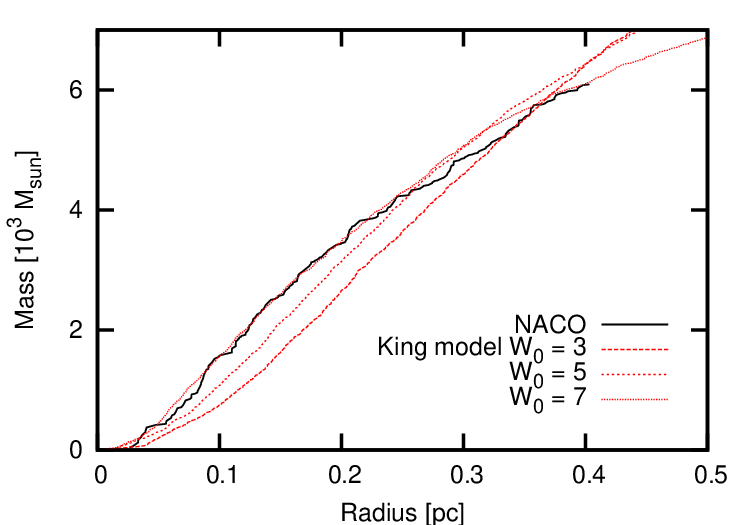}
\caption{The cumulative mass profile for stars with $m > 10\,\msun$.  }
\label{fig:obsnmofr}
\end{figure}


The field-of-view of the NACO data is such that only within a limited
radius all stars of the cluster can be seen. This radius is $0.4\,\pc$
(see left panel in Fig.~\ref{fig:bfmsnapshot} which shows an image of
all the $\sim1500$ cluster stars; the radius of $0.4\,\pc$ is
indicated by a circle centred on the centre of density). We therefore
also limit our sample to the $234$ massive stars within this
radius. The cumulative mass profile for these stars is shown in
Fig.~\ref{fig:obsnmofr}, where we also show the mass profiles for
three different King models \citep{King66}. All models have the same
virial radius $\rvir = 0.5\,\pc$ and are scaled in mass to
approximately fit the observed profile. The present-day profile of the
Arches cluster is best described by a King model with $W_0 = 7$.


\section{The model of the Arches cluster}
\label{sec:clustermodel}

In order to construct a model for the Arches cluster, we compare the
results of $N$-body simulations with the observational data described
above. The simulations start from a set of initial conditions with a
number of parameters that can be varied to find the best fitting
model. We have chosen the IMF, total mass, concentration, and size of
the cluster as the free parameters. Other parameters are fixed: we
assume an age of $2.5\,\myr$ \citep[see also][]{MHP08} and a distance
of $8\,\kpc$ to the cluster.



\subsection{Initial Mass Function}

It has been discussed whether the IMF of the Arches cluster deviates
from the norm, with a flattened distribution (for massive stars) with
respect to a Salpeter IMF. Determining the IMF of a star cluster from
its present-day mass function (PDMF) is not a straight-forward process
as the PDMF is the result of stellar evolution and dynamical
effects. Additional difficulties arise from observational limitations
like crowding. In the following, we will use the observed MF for the
comparison with our models. Since we do not take into account stellar
evolution in our models, the observed MF is determined using the
initial masses of stars (using the present-day masses of stars would
change the MF only for stars with $m>50\,\msun$). We also correct the
observed MF for the incompleteness of the data. The observed MF can
still differ from the underlying IMF due to the selection of stars
inside $R<0.4\,\pc$.

\begin{figure}
\includegraphics[width=84mm,angle=0]{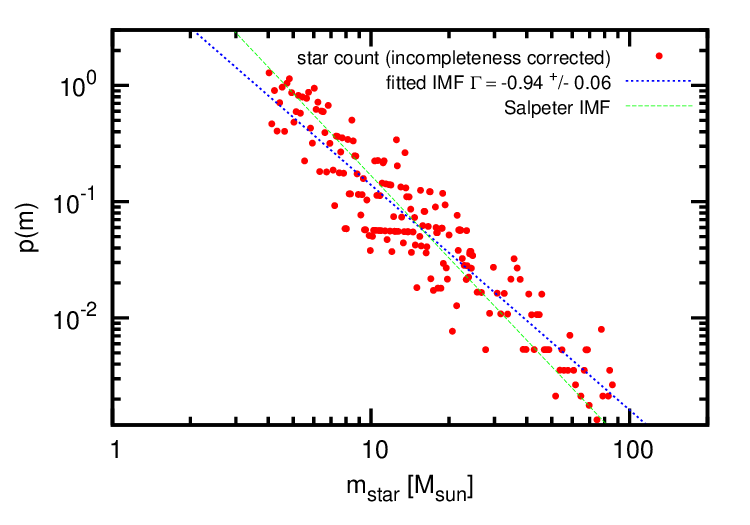}
\caption{The sampled probability distribution $p(m)$ of the observed
  MF of stars with $m > 4\,\msun$ in the Arches cluster (red
  dots). The dotted blue line is a fitted power-law MF and the green
  dashed line is a Salpeter MF. Data is from \citet{SBG05} and
  corrected for incompleteness.}
\label{fig:obsimf}
\end{figure}

In order to determine the slope of the observed MF, we use the sampled
probability distribution $p(m)$, which is defined as a stepwise
function through
\begin{equation}
\label{eq:pdist}
p(m) = \frac{2}{N} \left( m_{i+1} - m_{i-1}\right)^{-1} 
  \mathrm{for}\ m_{i-0.5}<m< m_{i+0.5}
\end{equation}
where $m_i$ are the sorted initial stellar masses and $m_{i\pm
  0.5}=0.5\cdot(m_i+m_{i\pm 1})$. In
Fig.~\ref{fig:obsimf}, we show $p(m_i)$ for the NACO data, where we correct
for the incompleteness of the data by dividing $p(m_i)$ by the
completeness fraction given in Fig.~\ref{fig:inc}.
The slope is derived by a least-square fit of power-law MF with two
free parameters (normalisation and slope). This allows a more
straight-forward fitting of the data than the commonly used mass
binning.  We find a shallow slope with $\Gamma = -0.94 \pm 0.15$ for
stars with $m > 4\,\msun$. \citet{Stolte03} has given an estimate of
$\pm 0.15$ for the total error, which we also use here. The formal
uncertainties of the fit are smaller (see Fig.~\ref{fig:obsimf}), but
these are not taking into account systematic uncertainties in
determining stellar masses.


\begin{figure}
\includegraphics[width=84mm,angle=0]{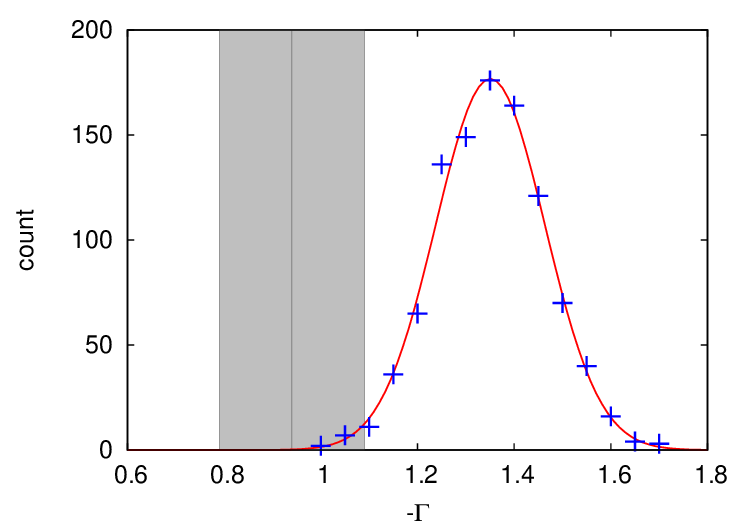}
\caption{The distribution of fitted $\Gamma$-values of model
  MFs. Shown are the results from the models (blue pluses) together
  with a fitted Gaussian distribution. The observed MF slope is
  indicated by the shaded area.}
\label{fig:simimf}
\end{figure}

The slope we find here is in agreement with what has been reported by
\citet{SBG05}, and significantly deviates from a standard Salpeter
MF.  As a test, we have created random realizations of the Salpeter
MF using $1\,\msun$ and $120\,\msun$ as the lower and upper mass
limit, respectively. The total number of models was 1000 and each
model consisted of 7500 stars which, on average, results in $320$
stars with $m > 10\,\msun$. We then fitted a power-law MF to each of
the models in the same way we fitted the observed MF, using only the
$\sim320$ massive stars. In Fig.~\ref{fig:simimf}, we show the
distribution of fitted $\Gamma$-values and from this distribution we
derive $\Gamma = -1.35 \pm 0.11$. The $\Gamma$-value derived from the
observations is indicated by the shaded box. Based on this, only a
small fraction of our models (nine) are consistent with the observed
MF. However, given the uncertainties in deriving the initial masses
of stars (which depend very much on the model for the rather unknown
mass loss rate) and since so far any effects from the dynamical
evolution of the cluster are not taken into account, we decided to use
the slope of the IMF as a free parameter and studied, how a different
IMF affects the present-day MF in the dynamically evolved and
mass-segregated cluster.  We also varied the lower mass limit of the
IMF to test for a possible truncation of the IMF at lower masses.

\subsection{Initial mass of the cluster}

The choice of an IMF is also important for determining the total mass
of the cluster. The present day mass of the cluster within
$R=0.4\,\pc$ is $\sim 1-2 \cdot 10^4\,\msun$
\citep{FNG02,SGM08,ESM09}. \citet{KFL00} estimate that Arches could
have lost about half its initial mass due to its dynamical evolution
in the Galactic centre, which would give an initial mass of $\sim3
\cdot 10^4\,\msun$. On the other hand, \citet{FNG02} calculated an
upper mass limit of $7 \cdot 10^4\,\msun$ for the cluster, based on
the virial theorem and the observed upper limit to the velocity
dispersion of $22\,\kms$ from radial velocities.

\begin{figure}
\includegraphics[width=84mm,angle=0]{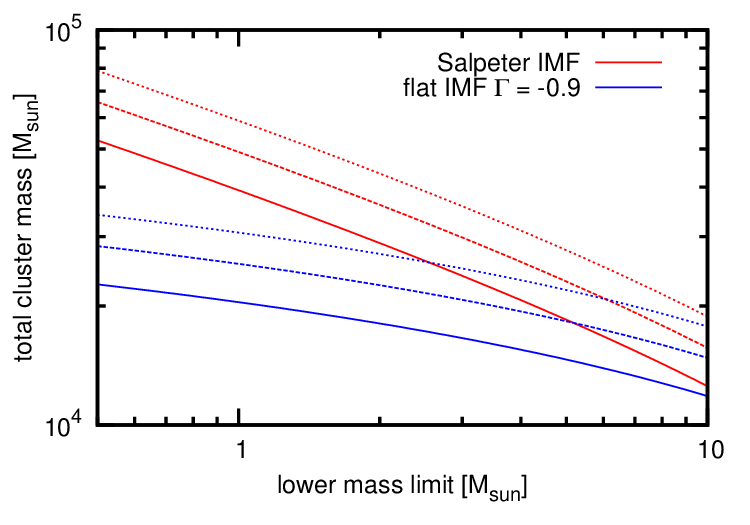}
\caption{The initial mass of the cluster as a function of the lower
  mass-limit of the IMF for different IMFs. The three lines show
  different normalisations.}
\label{fig:clustermass}
\end{figure}

In Fig.~\ref{fig:clustermass}, we show the initial cluster mass we
derive for different IMFs. For each IMF we used three different
normalisations such that the total number $\NMS$ of star more massive
than $20\,\msun$ is $200$, $250$, and $300$. We then varied the lower
mass limit of the IMF as it may be truncated in the Arches cluster
\citep{SBG05,KFK06}.

We find, that the total cluster mass can be initially as high as $\sim
8 \cdot 10^4\,\msun$ (Salpeter IMF and $\NMS = 300$) or even higher if
stars below $0.5\,\msun$ have formed in the cluster. For a flat IMF
($\Gamma = -0.9$), the total cluster mass does not depend much on the
lower cut-off. The average mass ($\NMS = 250$) is about $\sim3 \cdot
10^4\,\msun$ and a minimum initial mass of $\sim2 \cdot 10^4\,\msun$
for $\NMS=200$ is found, both in very good agreement with the findings
of \citet{KFL00,KFK06}.

\subsection{Size of the cluster}

The Arches is a very dense cluster with a central mass density of
$\sim2 \cdot 10^5\,\msun\,\pc^{-3}$ \citep{ESM09}. The tidal radius of
the Arches cluster is $\sim1\,\pc$ and we determine the core radius
with $r_\mathrm{core} = 0.25\,\pc$ using all stars and with
$r_\mathrm{core} = 0.14\,\pc$ for stars with $m >
10\,\msun$. Following \citet{CH85}, the core radius is defined
throughout this paper as the density-weighted average distance of
stars to the density centre.

The current concentration of the Arches cluster may be the result of
its dynamical evolution as the cluster is probably evolving towards
core collapse. We therefore use the initial concentration,
parameterised in the King model concentration parameter $W_0$, and
size of the cluster, namely its virial radius, as two more free
parameter. The virial radius $\rvir$ is defined as
\begin{equation}
\rvir = \frac{1}{2}\frac{GM^2}{\left|U\right|} 
\end{equation}
with the gravitational constant $G$, the mass $M$ and potential energy
$U$ of the cluster. Since $U$ cannot be observed, it may be more
practical to know that the virial radius is proportional to the half
mass radius.


\subsection{Summary of model parameters}

To summarise, the initial conditions for our cluster models are set by
the virial radius, slope and low-mass cutoff of the IMF,
concentration, and number of massive stars, which are varied
systematically (see Tab.~\ref{table:models} below and Sec. 4.1). 

\section{Setup and results of simulations}
\label{sec:simulations}

Our aim is to find a numerical model of the Arches cluster that can
explain the observed properties of the cluster. In this paper, we
neglect the effects of the Galactic potential as well as stellar
evolution. This allows us to perform a large number of simulations in
order to constrain some of the parameters of our models. The
simulations were carried out in the {\tt Starlab} environment using
the integrator {\tt kira} \citep{PZMH01}. GPUs, graphical processing
units, were used to accelerate the calculations via the {\tt Sapporo}
library \citep{GHPZ09}.

\subsection{Initial conditions}

\begin{table*}
\caption{List of models.}
\label{table:models}
\begin{tabular}{ccccccccc}
\hline
 Model & $W_0$ &IMF & $m_{\rm low}$ & $M_{\rm cluster}$ & $N_{\rm cluster}$ & $N(m>10\,\msun)$ & \multicolumn{2}{c}{parameter}\\
 &&& $[\msun]$ & $[10^3\msun]$ & $[10^3]$ && $R_{\rm vir} [\pc]$ & $N_\mathrm{MS}$ \\[2ex]
 IKW03F05 & 3 & flat & 0.5 & 22.9 & 6.9 & 423 & 0.1 -- 1.0 & 100 -- 300 \\
 IKW03F10 & 3 & flat & 1.0 & 20.5 & 3.7 & 421 & 0.1 -- 1.0 & 100 -- 300 \\
 IKW03S05 & 3 & Salpeter & 0.5 & 52.7 & 31.9 & 552 & 0.1 -- 1.0 & 100 -- 300 \\
 IKW03S10 & 3 & Salpeter & 1.0 & 39.7 & 12.5 & 552 & 0.1 -- 1.0 & 100 -- 300 \\
 IKW03S40 & 3 & Salpeter & 4.0 & 20.6 & 1.9 & 540 & 0.1 -- 1.0 & 100 -- 300 \\
 IKW05F05 & 5 & flat & 0.5 & 22.7 & 6.9 & 413 & 0.1 -- 1.0 & 100 -- 300 \\
 IKW05F10 & 5 & flat & 1.0 & 20.2 & 3.7 & 413 & 0.1 -- 1.0 & 100 -- 300 \\
 IKW05S10 & 5 & Salpeter & 1.0 & 39.0 & 12.5 & 545 & 0.1 -- 1.0 & 100 -- 300 \\
 IKW05S40 & 5 & Salpeter & 4.0 & 20.8 & 1.9 & 543 & 0.1 -- 1.0 & 100 -- 300 \\
 IKW07F05 & 7 & flat & 0.5 & 22.9 & 6.9 & 422 & 0.1 -- 1.0 & 100 -- 300 \\
 IKW07F10 & 7 & flat & 1.0 & 20.7 & 3.7 & 421 & 0.1 -- 1.0 & 100 -- 300 \\
 IKW07S10 & 7 & Salpeter & 1.0 & 39.2 & 12.5 & 537 & 0.1 -- 1.0 & 100 -- 300 \\
 IKW07S40 & 7 & Salpeter & 4.0 & 20.8 & 1.9 & 551 & 0.1 -- 1.0 & 100 -- 300 \\
\hline\\
\end{tabular}

\parbox{\textwidth}{\footnotesize Columns are: 1) model name; 2) the dimensionless King
concentration parameter $W_0$ 3) the IMF used, where flat and Salpeter
refer to power-law IMFs with $\Gamma=-0.9$ and $\Gamma=-1.35$,
respectively; 4) lower IMF mass limit in $\msun$; 5) total cluster
mass in $10^3\,\msun$; 6) total number of stars in the cluster in
$10^3$; 7) total number of massive stars ($m>10\,\msun$); 8)
additional model parameter (virial radius $\rvir$ and number of 
massive stars $\NMS$) and their ranges}

\end{table*}

We use King models \citep{King66} in virial equilibrium as our initial
cluster model. As the initial concentration of the cluster is not known,
it is possible that the cluster has evolved to its current compactness
from a less concentrated model. We therefore decided to use three
different values for the initial concentration parameter $W_0$. For
each concentration parameter we also tested both the Salpeter IMF and
an IMF with a flat slope of $\Gamma = -0.9$. Furthermore, we also
varied the lower mass limit of the IMF between $0.5\,\msun$ and
$4\,\msun$. In total, we used 13 different sets of initial conditions
for the cluster (see Tab.~\ref{table:models}).

In addition, we also varied two free parameters for each of these
models: the initial virial radius as $\rvir$ and the initial number
$\NMS$ of stars with $m>20\,\msun$. The latter is used to normalise
the total cluster mass and we used five different values for $\NMS$
between $100$ and $300$ in steps of $50$. This range covers the $127$
stars with $m>20\,\msun$ found in the NACO data and the $\sim 200$
stars reported by \citet{FNG02}, also taking into account that a
significant fraction of massive stars may no longer be bound to the
cluster or is not observed. The differences between the two data sets
can be explained by a different field-of-view ($25\arcsec$ squared in
the case of NACO as compared to $40\arcsec$ squared for NICMOS) and
the applied selection criteria to determine cluster membership.
Varying $\NMS$ increases the total number of models tested to $65$.

Our $N$-body simulations are in scale-free $N$-body units because we
neglect stellar evolution and the tidal field. In $N$-body units, the
gravitational constant $G$, the total mass of the system, and the
virial radius are all set to unity \citep{HM86}. The connection
between the scale-free $N$-body units and physical units is then given
by
\begin{equation}
  G = 1 \frac{\Ul^3}{\Um \Ut^2} = 0.0045 \frac{\pc^3}{\msun \myr^2},
\label{eq:nbodyunits}
\end{equation}
where $\Ul$, $\Um$, and $\Ut$ are the $N$-body units for length, mass,
and time, respectively. The mass unit $\Um$ is naturally given by the
total mass of the Arches cluster $M_\mathrm{Arches}$. A choice of
$\rvir$ defines $\Ul$, which in turn determines $\Ut$ via
Eq.~\ref{eq:nbodyunits}. The age of the cluster is $t=2.5\,\myr$ or
according to Eq.~\ref{eq:nbodyunits}
\begin{equation}
  t = \frac{2.5\,\myr}{\Ut} = 2.5 \cdot \sqrt{ 0.0045 \frac{M_\mathrm{Arches}}{\msun} \left(\frac{\rvir}{\pc}\right)^{-3}}
\label{eq:nbtime}
\end{equation}
in the dimensionless $N$-body time unit.

The virial radius was varied between $0.1$ and $1\,\pc$ in steps of
$0.05\,\pc$. However, instead of increasing the number of models by
another factor of 19, we make use of the scale-free nature of our
simulations. For each value of $\rvir$, the cluster has to be evolved
to a different $N$-body time unit given by Eq.~\ref{eq:nbtime} to
reach an age of $2.5\,\myr$ in physical units. The $N$-body times to
match the desired values of $\rvir$ can be computed at the start of
the simulation, and then snapshots are written at these $N$-body times
during these simulations. Each of the snapshots corresponds to a
snapshot of the Arches cluster at $t=2.5\,\myr$ with a different
$\rvir$. 


Each of the $65$ models was created ten times with a different random
realization, so that a total of $650$ simulations were
performed. However, once the simulations were finished, we averaged
the global properties of each model before comparing them with the
observations. We also took into account the incompleteness of the
observations by randomly removing a few stars from our models
according to the incompleteness tests depicted in
Fig.~\ref{fig:inc}. However, this has only little effect on the
overall results as we already constrained our data sample to stars
that are almost complete in the observations. In the end, we compared
more than 1,200 simulation snapshots with the observations.


\subsection{Comparing models and observation}

\begin{figure*}
\includegraphics[width=84mm,angle=0]{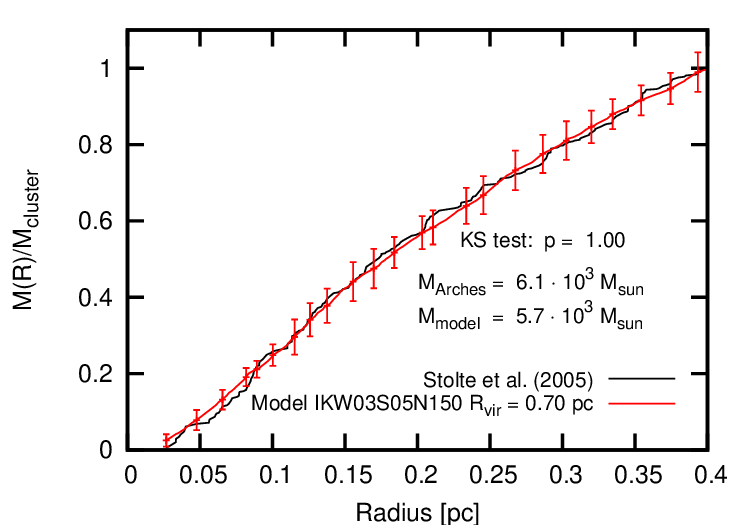}
\includegraphics[width=84mm,angle=0]{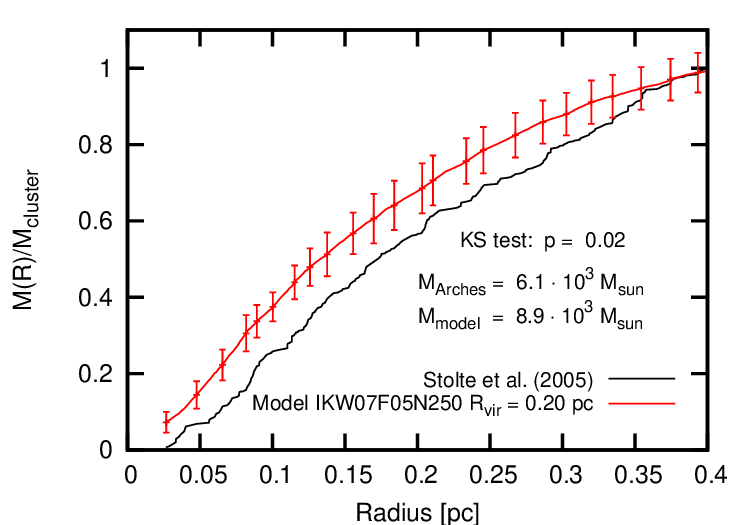}
\caption{Comparing the normalised cumulative mass profile of model and
  observation. The red line shows the model data, with error bars
  indicating the standard deviation from averaging over ten random
  realization of the same model. The black line the observation and
  $M_\mathrm{Arches}$ is the total observed cluster mass for
  $m>10\,\msun$ and $R\leq0.4\,\pc$. Two different models are shown.}
\label{fig:modvsobs_mprof}
\end{figure*}


\begin{figure*}
\includegraphics[width=84mm,angle=0]{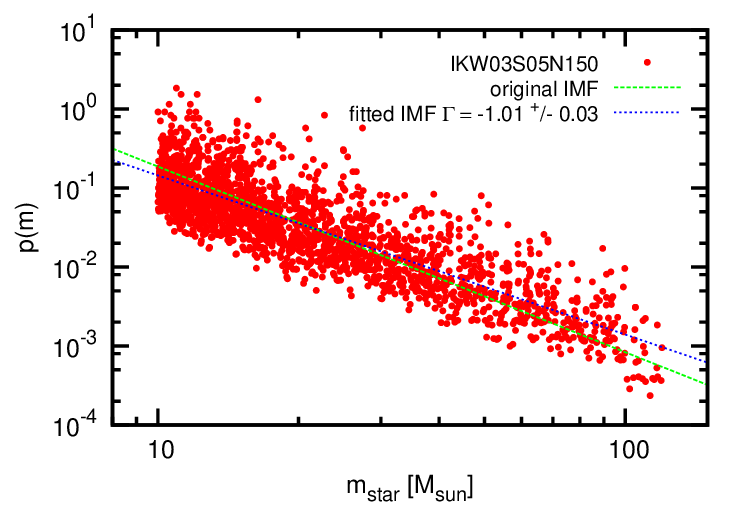}
\includegraphics[width=84mm,angle=0]{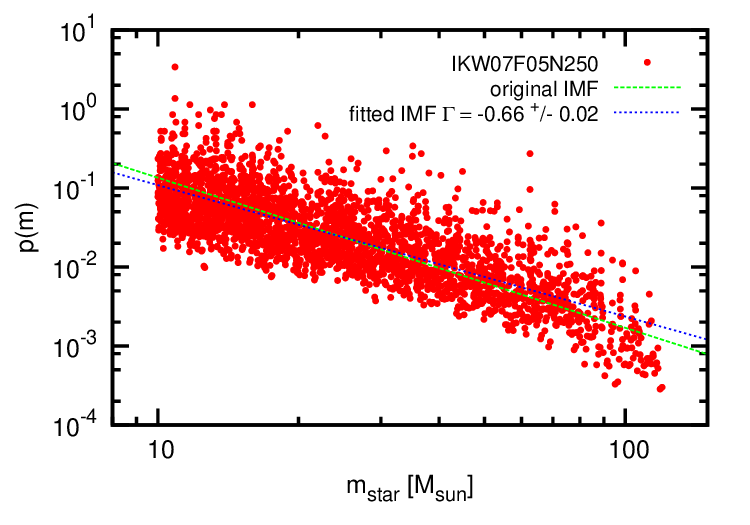}
\caption{Determination of the model MF for $R\leq0.4\,\pc$ The red
  points show the sampled probability distribution $p(m)$ for model
  data, adding together ten random realization of the same model. The
  dotted blue line indicates a least-square power-law fit to the data
  and the slope $\Gamma$ can be compared to the observed MF slope in
  Fig~\ref{fig:obsimf}. The dashed green line shows the original IMF
  that was used to setup the ICs for the model. Two different models
  are shown. }
\label{fig:modvsobs_imf}
\end{figure*}

In order to compare the simulation snapshots with the NACO data, we
first computed a cumulative mass profile for each parameter set. In
this process, we averaged the ten different random realizations and
obtained a single mass profile. We only selected stars with $m>
10\,\msun$ and within $R\leq0.4\,\pc$ and we also normalised the
profile to the total mass inside this radius (in the following
comparisons we compare the shape of the profile and the total mass
separately). Two of the resulting profiles are shown in
Fig.~\ref{fig:modvsobs_mprof} (red line with error bars) in comparison
with the observed profile (black line).

In the following we define a number of fitness parameters $f$ which
describe the quality of the fit of the model to the observation. These
parameters are defined such that the values range from zero to
unity. A value close to unity describes a good fit.

The mass profile fitness parameter $\fCMP$ is used to quantify how
well the cumulative mass profile of the model fits the
observations. We employ a two-sample KS test \citep[see][]{PTV92} to
compare random samples of the two mass profiles. The KS-test then
returns the probability $p$, that the two samples are drawn from the
same $M(R)$-distribution. In Fig.~\ref{fig:modvsobs_mprof}, the
$p$-value is given for two different models and it can be seen that a
good fit results in a high $p$-value as expected. In the following, we
will use
\begin{equation}
  \fCMP = p,
\end{equation}
where a value of $\fCMP$ close to unity describes an excellent fit. 

We also compare the total mass of stars with $m>10\,\msun$ inside of
$0.4\,\pc$. We define the mass fitness parameter 
\begin{equation}
\fM = \exp\left( -\frac{1}{2}
  \left[\frac{M_\mathrm{Arches} - M_\mathrm{model}}{\Delta M_\mathrm{Arches}}\right]^2 \right)
\end{equation}
as an estimator for how well the observed total mass of the cluster is
reproduced in the model. \revised{From the NACO data, we determine the
  observed, incompleteness corrected cluster mass $M_\mathrm{Arches} =
  6.1\cdot 10^3\msun$. The uncertainty in the total mass is estimated
  with $\Delta M_\mathrm{Arches} = 1\cdot 10^3\msun$ which accounts
  for uncertainties in individual stellar masses, cluster age, and
  stellar evolution models.} With the above definition, $\fM = 1$
identifies a perfect match and becomes smaller the more
$M_\mathrm{model}$ deviates from $M_\mathrm{Arches}$.


In Fig.~\ref{fig:modvsobs_imf}, we show the comparison of the observed
MF for the Arches cluster with our models. The sampled probability
distribution (Eq.~\ref{eq:pdist}) of the model (red dots) is fitted by
a power-law MF (dotted blue line) to determine the slope $\Gamma$. The
resulting $\Gamma$ can be compared with the observed MF
(Fig.~\ref{fig:obsimf}). Two different models are shown, which started
out with a Salpeter IMF (left panel) and a flat IMF (right panel). The
IMF used for the initial conditions of the model is also plotted in
each panel (dashed green line). Because we only take stars within
$0.4\,\pc$ into account, the measured (observed) MF is flattened
significantly by $0.2-0.3$ from the original IMF due to dynamical
evolution. Similar to $\fCMP$, we also define a MF fitness parameter
$\fIMF$ as
\begin{equation}
\fIMF = p
\end{equation}
where $p$ again is the KS-test probability that the observed MF and
the model MF are from the same distribution.

\begin{figure*}
\includegraphics[width=84mm,angle=0]{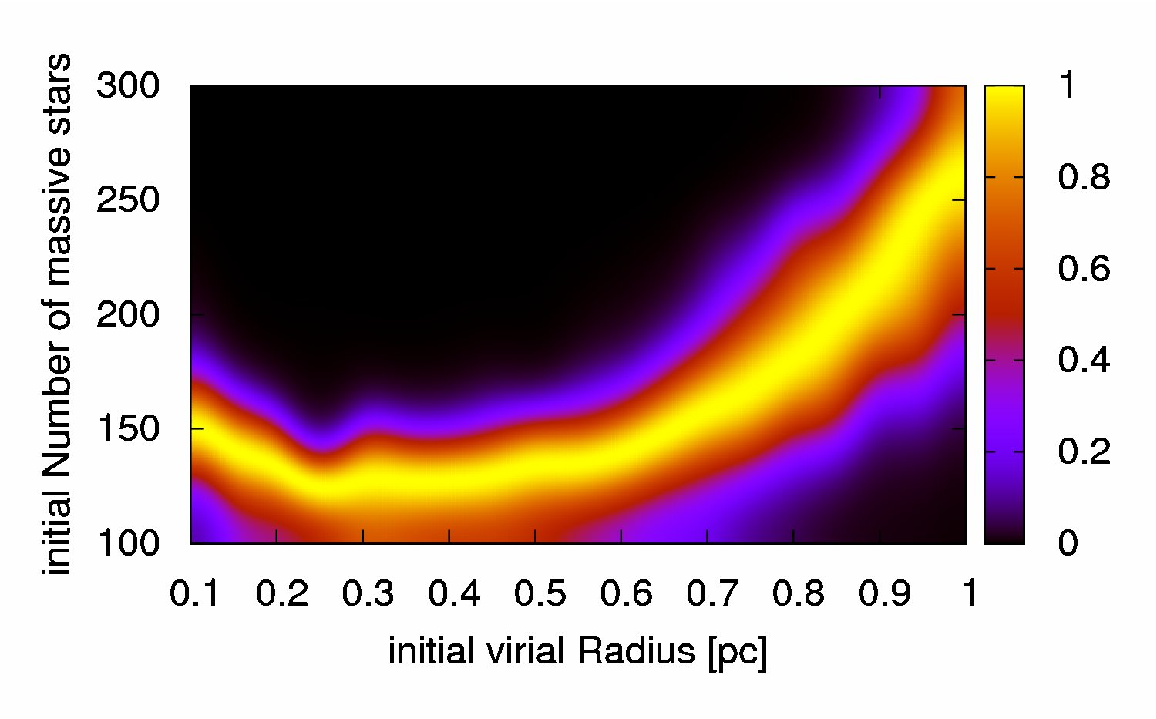}
\includegraphics[width=84mm,angle=0]{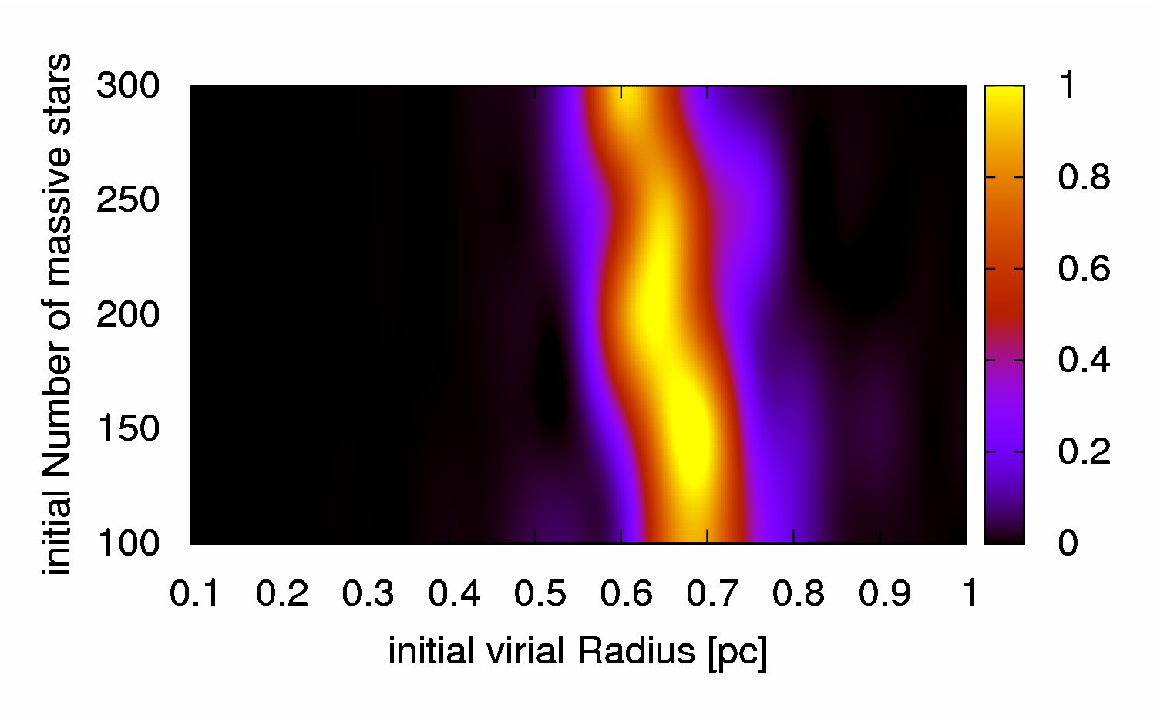}
\includegraphics[width=84mm,angle=0]{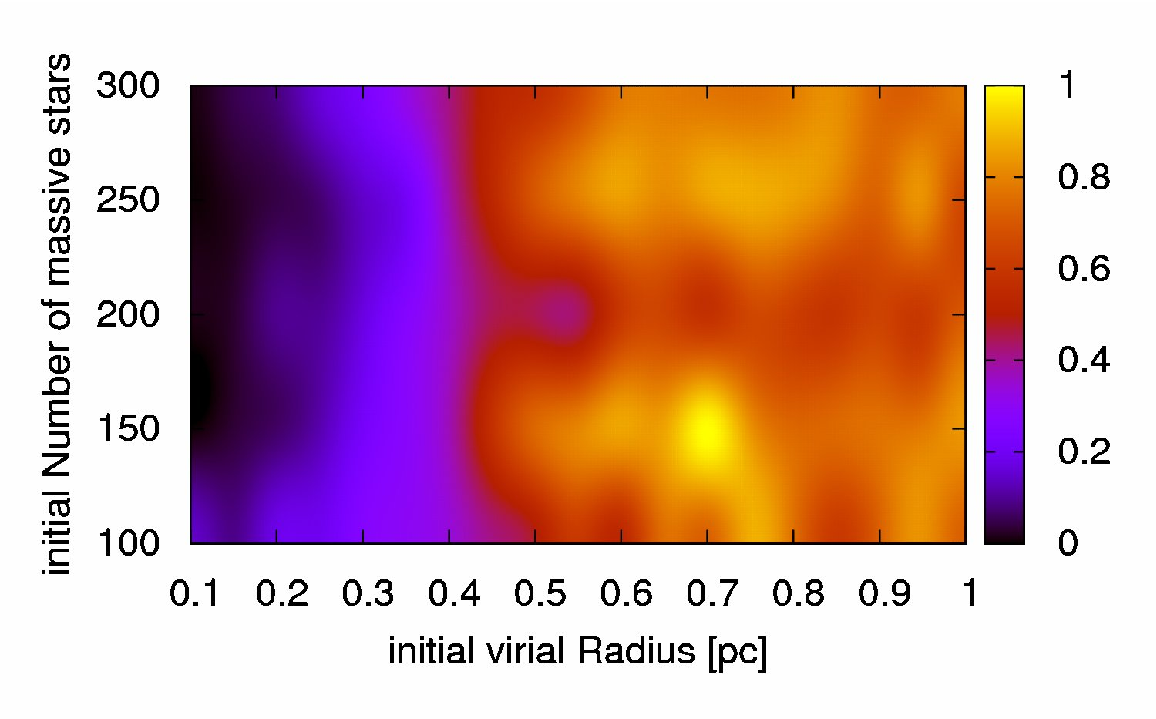}
\includegraphics[width=84mm,angle=0]{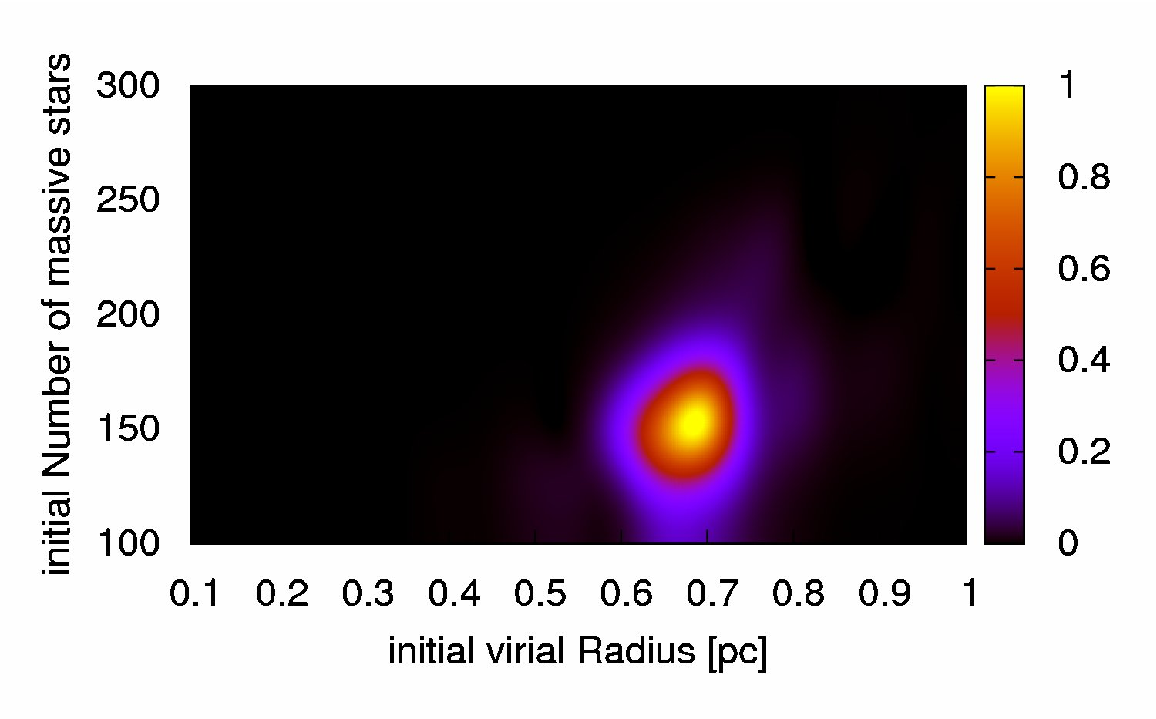}
\caption{Quality of fit to observations for varying model parameters
  $\rvir$ and $\NMS$. Data from model IKW03S05 (isolated King model
  with $W_0 = 3$ and a Salpeter IMF with $m_\mathrm{low} =
  0.5\,\msun$). Panels show the fit to the cluster mass (top left),
  cumulative mass profile (top right), IMF (bottom left) and product
  of all three fits (bottom right). A fitness parameter close or equal
  to unity indicates the best fit models and in the final panel an
  overall best fit model can be clearly identified.}
\label{fig:modelfit}
\end{figure*}

The quality of the fit for model IKW03S05 is presented in
Fig.~\ref{fig:modelfit}, which shows how $\fCMP$, $\fM$, and
$\fIMF$ vary for the model parameters $\rvir$ and
$N_\mathrm{MS}$. The cluster mass is best fitted with $\NMS \approx
150$ and almost independent of $\rvir$. Only models with $\rvir \ge
0.8$ require larger values of $\NMS$. This can be explained as
follows: models with small $\rvir$ evolve faster dynamically and these
models have gone through core collapse already at an age of
$2.5\,\myr$. As a result, the initial dependence of $\NMS$ on $\rvir$,
which can still be seen for larger values of $\rvir$, is lost.

The top right panel in Fig.~\ref{fig:modelfit} shows which models have
the best fitting mass profiles. In this case, the best models are
found within a narrow range of $\rvir$-values with no dependence on
$\NMS$. Models with $\rvir \le 0.5\,\pc$ are, in comparison with the
observations, too concentrated (see also
Fig.~\ref{fig:modvsobs_mprof}) whereas $\rvir > 0.8\,\pc$ results in a
too shallow mass profile.

The MF depends not as strongly on the model parameters (bottom left
panel) though the general trend is, that the best fits are found for
values of $\rvir$ larger than $0.4\pc$. In the last panel, we show the
combined fitness $\fall$ which is defined as
\begin{equation}
  \fall = \fCMP \cdot \fM \cdot \fIMF.
\end{equation}
As a result, the best fitting model in the model series IKW03S05 can
be clearly identified and it has an initial virial radius $\rvir =
0.70\,\pc$ and $\NMS = 150$. The number of stars with $m>10\,\msun$
and inside $R<0.4\,\pc$ in this particular model is $208 \pm 12$
(averaged over ten random realizations), compared to $234$ in the
observed sample.

\begin{figure*}
\includegraphics[width=84mm,angle=0]{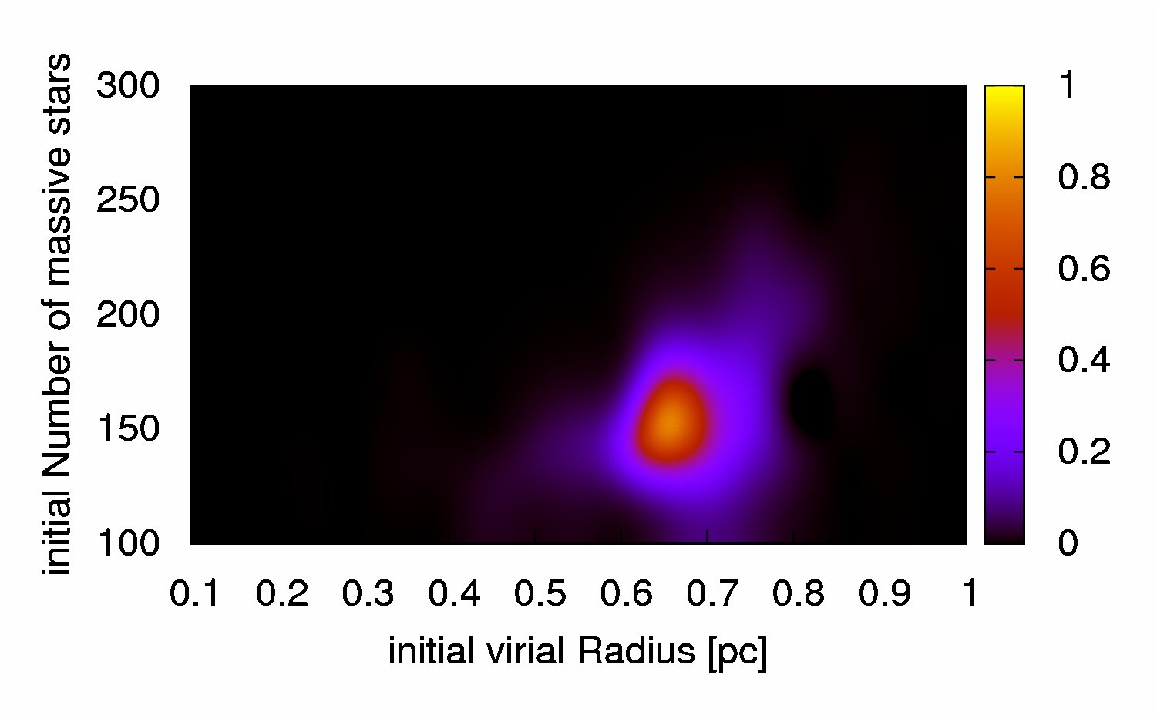}
\includegraphics[width=84mm,angle=0]{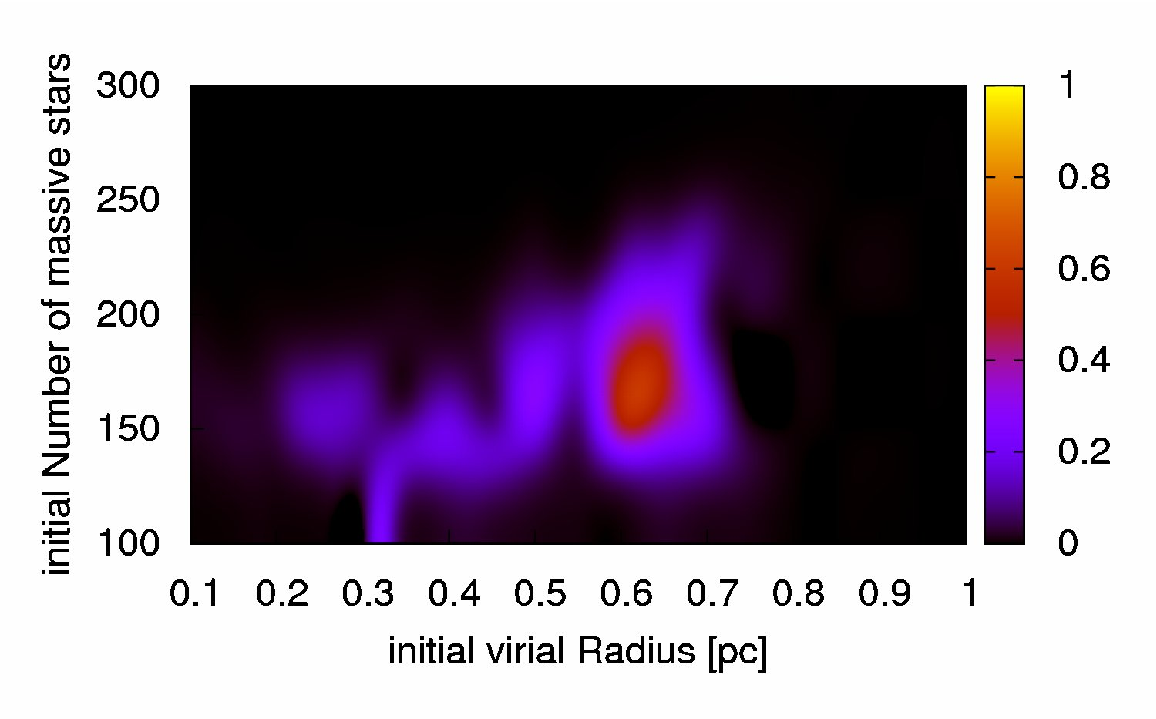}
\includegraphics[width=84mm,angle=0]{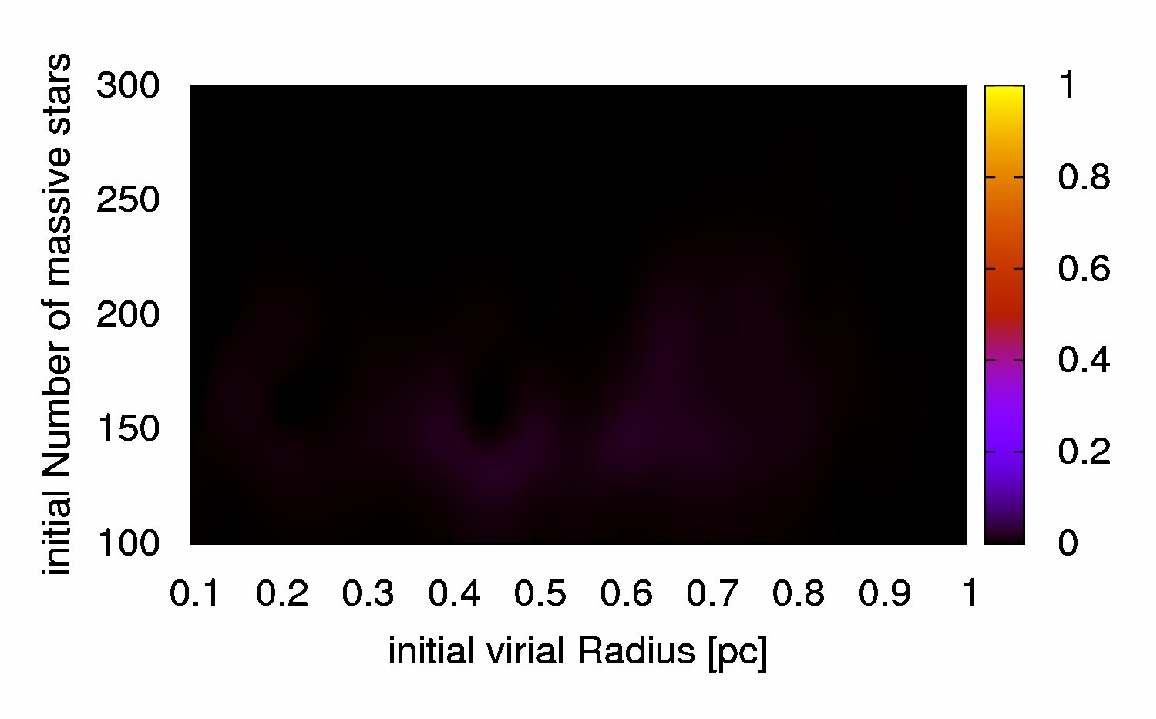}
\includegraphics[width=84mm,angle=0]{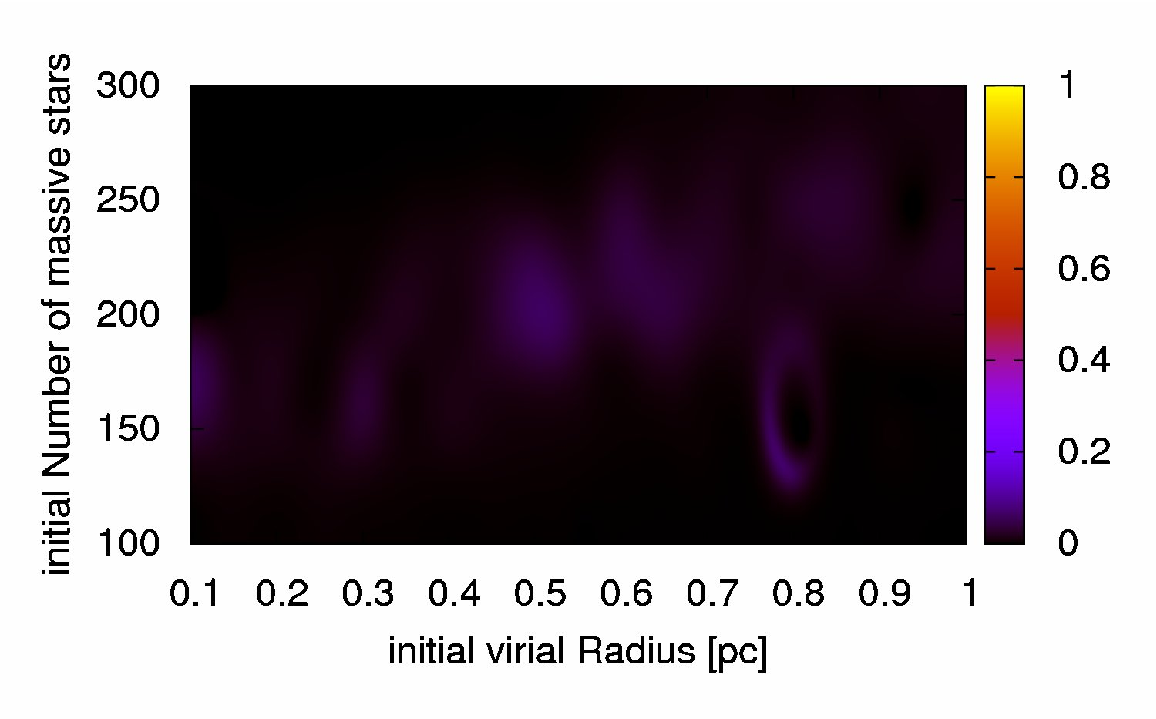}
\caption{Quality of fit to observations for different models and
  varying model parameters $\rvir$ and $\NMS$. The combined fitness
  $\fall$ is shown for models IKW03S10 (top left), IKW03S40 (top
  right), IKW03F05 (bottom left), and IKW07F10 (bottom right).}
\label{fig:modelcmp}
\end{figure*}

In Fig.~\ref{fig:modelcmp} we compare $\fall$ for a number of
different model series. The two panels at the top show the results for
the models IKW03S10 and IKW03S40. The only difference between these
two models and also to model IKW03S05 shown in Fig.~\ref{fig:modelfit}
is the lower mass limit of the IMF, which is $1.0$, $4.0$, and
$0.5\,\msun$, respectively. In each of the three models the best fit
is in a similar regime of the parameter space, with $\rvir =
0.6-0.7\,\pc$ and $\NMS \approx 150$. In all other model series that
have an acceptable fit, the best fitting models also lie within a
small area of the parameter space.

The two bottom panels show results from models starting initially with
a flat IMF. These models produce no acceptable fit for any of our
tested parameter combinations. The reason is that the slope of the IMF
is always flattened by the dynamical evolution of the cluster (see
also Fig.~\ref{fig:modvsobs_imf}), so that models starting with the
observed slope cannot produce a good fit. This result is independent
of the initial concentration of the cluster and the lower cut-off mass
of the IMF.


\begin{table}
\caption{Best fitting models.}
\label{table:bestfit}
\begin{tabular}{ccccc}
\hline
 Model & \multicolumn{2}{c}{parameter} & \multicolumn{2}{c}{$\fall$} \\
       & $\rvir [\pc]$ & $N_\mathrm{MS}$ & $m>10\,\msun$ & $m>4\,\msun$ \\[2ex]
 IKW03F05 & 0.6  & 150 & 0.02 & 0.03 \\
 IKW03F10 & 0.35 & 150 & 0.06 & 0.03 \\
 IKW03S05 & 0.7  & 150 & 0.94 & 0.48 \\
 IKW03S10 & 0.65 & 150 & 0.78 & 0.52 \\
 IKW03S40 & 0.6  & 150 & 0.44 & 0.36 \\
 IKW05F05 & 0.45 & 150 & 0.04 & 0.07 \\
 IKW05F10 & 0.25 & 150 & 0.05 & 0.10 \\
 IKW05S10 & 0.8  & 200 & 0.65 & 0.44 \\
 IKW05S40 & 0.8  & 200 & 0.64 & 0.47 \\
 IKW07F05 & 0.75 & 250 & 0.05 & 0.04 \\
 IKW07F10 & 0.5  & 200 & 0.06 & 0.04 \\
 IKW07S10 & 1.0  & 200 & 0.55 & 0.36 \\
 IKW07S40 & 0.85 & 250 & 0.62 & 0.40 \\
\hline\\
\end{tabular}

\footnotesize Columns are: 1) model name; 2) values of additional
model parameter; 3) and 4) $\fall$ comparing stars with
$m>10\,\msun$ and $m>4\,\msun$

\end{table}

We have summarised the results in Tab.~\ref{table:bestfit} where for
each model series the best values for $\rvir$ and $\NMS$ are given
together with the corresponding value of $\fall$. We also repeated the
whole analysis using all stars down to $4\,\msun$ in the
comparison. In both cases, models with the highest values of $\fall$
all have a Salpeter IMF, \revised{though the comparison down to
  $4\,\msun$ generally yields lower values for $\fall$. This is not
  surprising as the uncertainties due to incompleteness and field
  contamination increase towards lower masses.}  \revised{It should
  also be noted, that the overall best fit model (IKW03S05) clearly
  stands out with $\fall = 0.94$, when the next best model has
  $\fall=0.78$ (both models are different only in the lower mass
  limit).}

The best fit models with a Salpeter IMF also have similar value of
$\rvir$ and $\NMS$. We computed the average of these values weighted
with $\fall$ which results in $\rvir = 0.77 \pm 0.12\,\pc$ and $\NMS =
183 \pm 35$ for the comparison using stars with $m>10\,\msun$. The
initial mass of the cluster depends on the lower mass limit of the
IMF. Our best model (and the favoured solution) has a lower mass limit
$m_\mathrm{lim}$ of $0.5\msun$, however models with a higher $m_{lim}$
also produce acceptable fits. Therefore, $m_{lim}$ is not well
constrained by our models, \revised{which is to be expected since we
  only compare stars with $m>10\,\msun$. In addition, observations are
  also limited by incompleteness and field contamination at lower
  masses ($m<4\,msun$ for the NACO data)}. Based on our results for
stars with $m>10\,\msun$, we get $M = \left(4.9\pm0.8\right)\cdot
10^4\,\msun$, $M = \left(3.6\pm0.6\right)\cdot 10^4\,\msun$, and $M =
\left(1.9\pm0.3\right)\cdot 10^4\,\msun$ for lower mass limits of the
IMF of $0.5$, $1$, $4\,\msun$, respectively.

\section{Discussion}
\label{sec:conclusions}

\begin{figure*}
\includegraphics[width=84mm,angle=0]{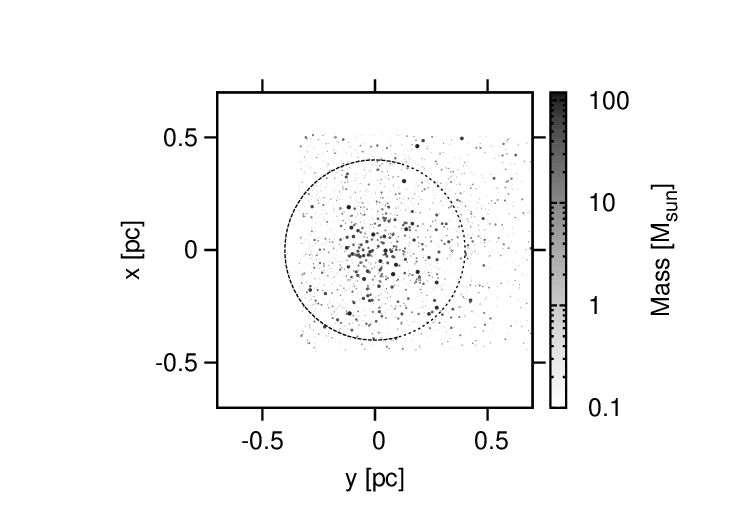}
\includegraphics[width=84mm,angle=0]{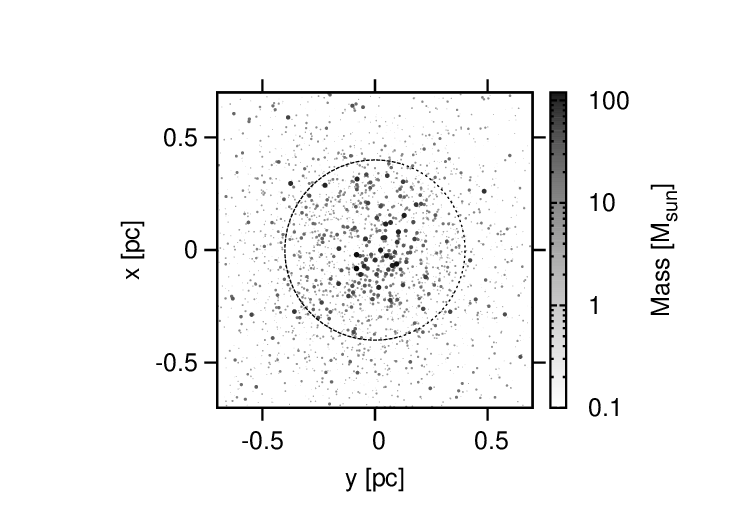}
\caption{The observed cluster (left) and a snapshot of one of the best
  fitting models (IKW03S05, right) in comparison. The images are
  centred on the centre of density and the circles indicate a radius
  of $0.4\,\pc$. Gray-scale and point size represent stellar
  masses. In the right panel, stars have been removed randomly to
  mimic incompleteness.}
\label{fig:bfmsnapshot}
\end{figure*}

A number of models in Tab.~\ref{table:bestfit} can be considered best
fit models, with the overall best fit being model IKW03S05 with $\rvir
= 0.70$ and $\NMS = 150$. A snapshot of this model at $t=2.5\,\myr$ is
shown in Fig.~\ref{fig:bfmsnapshot}. Stellar masses are indicated by
the point size and gray-scale and the centre of density of the cluster
is located at the origin. The dashed circle has a radius of
$0.4\,\pc$. In the left panel, the NACO data is plotted for comparison
in the same way as the simulation data in the right panel. In the
latter, stars have been randomly removed to mimic incompleteness. The
probability that a star is removed is given by a function fitted to
the data shown in Fig.~\ref{fig:inc} and depends on mass and position
of the star.

At $2.5\,\myr$, the cluster in model IKW03S05 has a total mass of
$\sim1.8\cdot10^4\,\msun$ inside a radius of $0.4\,\pc$ and twice that
mass inside the tidal radius of $1\,\pc$.  Inside of $0.4\,\pc$, about
$3\cdot10^3\,\msun$ of the total mass are in stars below $1\,\msun$
($9\cdot10^3\,\msun$ for $R<1\,\pc$). \citet{ESM09} determined the
cluster mass with $(2\pm0.6)\cdot10^4\,\msun$ which is in agreement
with out results. The models suggest that about half of the cluster's
mass is located in an annulus with $0.4<R<1\,\pc$. This result is in
good agreement with the previous findings of \citet{PZMM02}.

Generally, only models with the Salpeter IMF produce acceptable fits
and we therefore conclude that the IMF in the Arches cluster is
consistent with having a normal Salpeter slope despite the extreme
environment in which the cluster is formed. However, we cannot rule
out a turn-over as suggested by \citep{SBG05} as our results are not
very sensitive to the lower mass limit of the IMF. Our best fitting
models can have any lower mass limit in the the range of $0.5 -
4.0\,\msun$ that we investigated, though the overall best fit has a
lower limit of $0.5\msun$.

\begin{figure}
\includegraphics[width=84mm,angle=0]{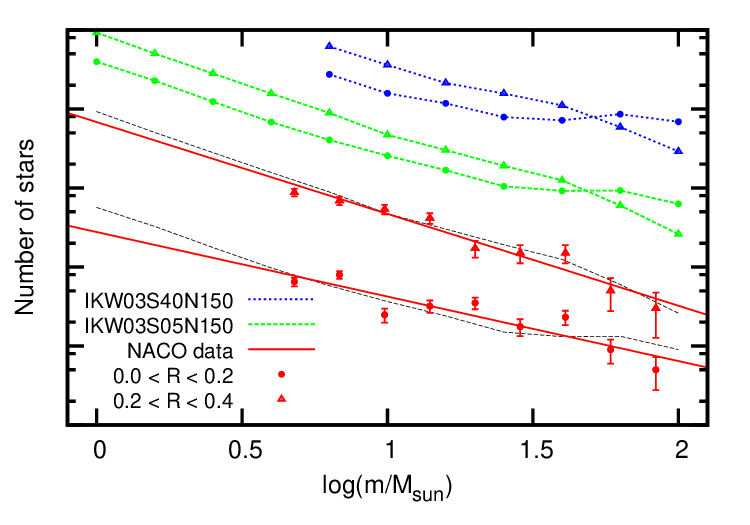}
\caption{Comparing the observed MF in two radial bins. The full red
  lines are fitted power-law MF for the NACO data. The dashed green
  and the dotted blue lines connect the data points of the best fit
  models from the model series IKW03S05 and IKW03S40,
  respectively. The circles and triangles indicate data points from
  different radial bins and shifts of one or two dex have been applied
  to clarify the plot. For a better comparison, model IKW03S05 is also
  plotted without shifting (thin black line). }
\label{fig:bfmimf}
\end{figure}

In Fig.~\ref{fig:bfmimf}, we compare the observed MF in two radial
bins, namely $R < 0.2\pc$ and $0.2 < R < 0.4\pc$ (the same as used by
\citet{SBG05} and \citet{ESM09}) with the observed model MF after
$2.5\,\myr$. A bin size of $\Delta \log(m/\msun) = 0.2$ was used to
obtain the individual star counts and the data from our best fit
models was also averaged over the ten different realizations
simulated. Two of the best fit models, IKW03S05 and IKW03S40, and the
NACO are compared. For each data set, the circles and triangles mark
the data points for the inner and outer radial bin, respectively. The
full red lines show a fitted power-law MF for the NACO data. The
dashed green and dotted blues lines connect the model data points.

The NACO data is fitted well by a single power-law and no break in the
MF can be seen. We find slopes of $\Gamma = -0.8\pm0.1$ and $\Gamma =
-1.2\pm0.1$ for the inner and outer radial bins, respectively. These
slopes are slightly shallower than the slopes recently reported by
\citet{ESM09}. The two model MFs for the outer radial bin can also be
described by a single power-law with much the same slope. The model
MFs in the inner radial bin are however noticeably flattened for
massive stars ($m \gap 10\,\msun$). This behaviour can be best fitted
by a broken power-law MF with a turning point at $m\approx 20\msun$
and a shallow slope of $\Gamma = -0.3\pm0.1$ at the high-mass end. In
a related study, \citet{PZGC07} also found a broken power-law MF,
though in a radial bin similar to our outer bin (no data for an inner
radial bin was shown \citet{PZGC07}). The turning point in their
favoured model was at $5-6\msun$, however. They also noticed, that the
break mass depended on the lower mass limit of the IMF which is not
the case for our two best fit models IKW03S05 and IKW03S40. Both
models have the same break mass but different lower mass limits ($0.5$
and $4\msun$, respectively). \revised{The reason for these differences
  most likely is the different scaling used: \citet{PZGC07} scale
  their models by the time scale $t_\mathrm{cc}$ for core collapse,
  while here we scale to physical units. Therefore, our models have
  evolved for different fractions of $t_\mathrm{cc}$ which results in
  different break masses (see also Fig.~2 in \citet{PZGC07}).}

In the Arches cluster, a possible turning point of the central MF
could be indicated by the location of $5-6\msun$-bump
\citep{KFK06}. The turn-over reported by \citet{SBG05} is also at this
mass, however, models show an increase in the MF slope instead of the
observed low-mass depletion. Generally, the observational data (both
NACO and the data shown in \citet{KFK06}) can be well described by a
single power-law, \revised{whereas our models are better fitted by a
  broken power-law for the inner MF. Yet, the present-day MF from the
  observations and the model are still consistent with each other
  within the uncertainties of the observational data, as shown by the
  direct comparison between the NACO data and model IKW03S05 (thin
  black line in Fig.~\ref{fig:bfmimf}). Alternatively, a possible
  difference between the models and the observationally derived MFs}
could result from uncertainties in the stellar evolution models and in
the incompleteness correction. The models, on the other hand, do not
include the effects of stellar evolution and the tidal field. Another
explanation could be primordial mass segregation \citep{CGU09}, which
is not included in the present simulations.

\begin{figure}
\includegraphics[width=84mm,angle=0]{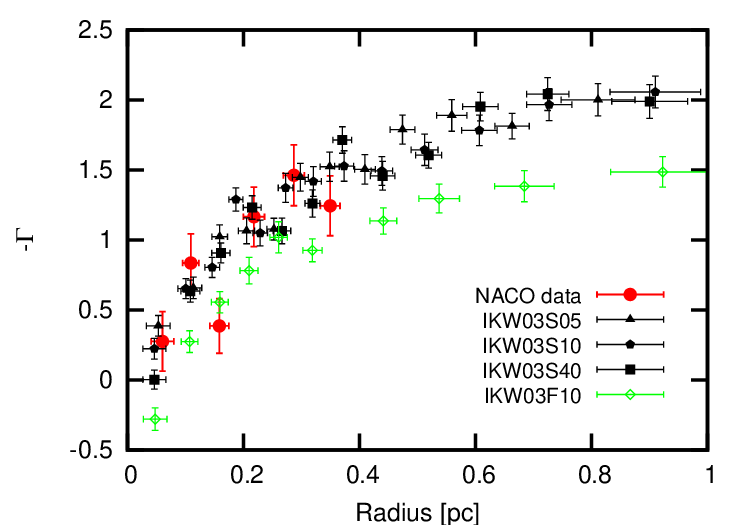}
\caption{The slope $\Gamma$ of the observed MF as a function of
  radius. The NACO data is shown (red circles and error bars) in
  comparison to four best fit models (three with Salpeter IMF and one
  with a flat IMF).}
\label{fig:gammavsr}
\end{figure}

In Fig.~\ref{fig:gammavsr}, we compare the slope $\Gamma$ of the
observed MF that we derived from the NACO data and our best fit
models. For each data set, we binned an equal number of stars into
radial bins and determined $\Gamma$ as described before. Three of the
models have a Salpeter IMF (black filled symbols) with different lower
mass limits. The fourth model started with a flat IMF (green open
symbol). In all four models and also the NACO data, the same
flattening of the MF towards the centre can be seen. \revised{However,
  in comparison, the three models with a Salpeter IMF are in better
  agreement with the NACO data than the model with the flat IMF.  The
  MF slopes of the latter are shifted by $\sim 0.4$ towards more
  positive $\Gamma$-values for $R<0.4\pc$ and even more at larger
  radii. The shift in the inner $0.4\pc$ is comparable to the initial
  difference between the Salpeter and the flat IMF.} The flattening of
the MF in the cluster core has been reported in all the previous
observational studies, though with varying $\Gamma$-values determined
for the slope. Our best fit models are in very good agreement,
however, with the latest results from \citet{ESM09} and also the
results from \citet{PZGC07}. \revised{Our models predict a much
  steeper MF (steeper than Salpeter) for massive stars ($m>10\msun$)
  at radii $R>0.4\pc$. It would therefore be very interesting to
  determine the MF slope at these larger radii. Unfortunately, it
  becomes very challenging to distinguish cluster and field stars
  beyond $0.4\pc$.}

Dynamical Mass segregation combined with a radially limited selection
of stars make the observed MF appear shallower than the global MF
truly is. Stars with $m>10\,\msun$ can be found in our models as far
as $10\,\pc$ from the cluster centre. However, we find that in order
to measure the correct slope of the MF, we only need the stars inside
the tidal radius of $1\,\pc$.

No turn-over at the low mass end of the IMF can be seen. This means
that the turn-over seen by \citet{SBG05} cannot be explained by mass
segregation (unlike the flat IMF at the high-mass end). Deriving the
IMF from our simulation data is not hampered by incompleteness,
selection effects, and mass determination from observed
luminosities. All these effects may bias the observed star counts to
produce a turn-over, however, the turn-over appears at $\sim6\,\msun$
where the data is still 50\% complete. So, either the IMF is indeed
truncated or another effect has to be considered. One possible
explanation is that tidal stripping preferably removes low-mass stars
from the cluster. This effect is not included in our current
simulations but will be tested in a follow-up paper. Alternatively,
\citet{ESM09} pointed out that local variations in extinction can
account for (some of) the flattening of the observed MF. The same
effect could leave lower-mass stars to be undetected
preferentially. Then the completeness function, which only takes into
account crowding and sensitivity effect, would not be sufficient to
correct the low-mass end of the IMF.

\begin{figure}
\includegraphics[width=84mm,angle=0]{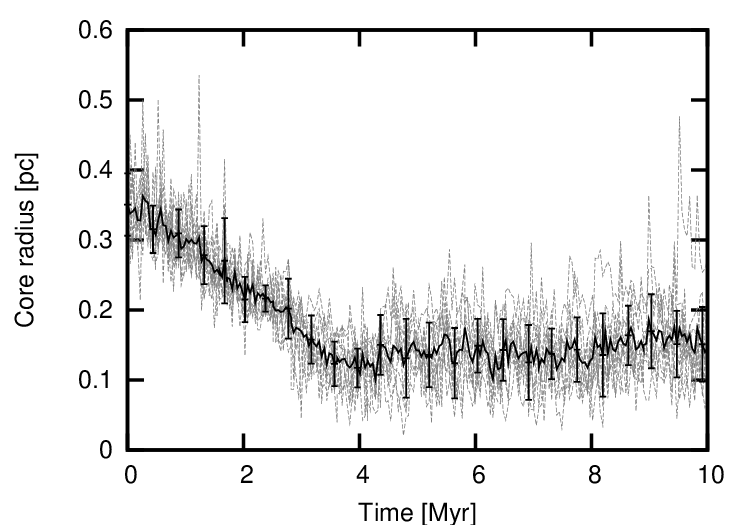}
\caption{The core radius evolution with time of model IKW03S05 with
  $\rvir=0.7\,\pc$ and $\NMS=150$.  The black line with errorbars
  shows the average of ten random realizations of the same model (grey
  lines). Core collapse happens at $\sim4\,\myr$. }
\label{fig:rcore}
\end{figure}

In Fig.~\ref{fig:rcore} we show the core radius evolution with time
for one of our best fit models (IKW03S05). At its current age of
$2.5\,\myr$ the cluster is more than halfway to core collapse which
will occur at $\sim4\,\myr$ and a little more dynamically evolved than
the favoured model in \citet{PZGC07}.

\section{Summary}
\label{sec:summary}

We have performed a large number of $N$-body simulations in order to
find the best fitting model for the Arches cluster. The available
observational data has been used to constrain the free parameters in
our model. In a systematic analysis, we compared the total mass, the
cumulative mass profile, and the present-day MF and defined fitness
parameters for each of the three observables.

The main conclusion from our analysis is that the Arches cluster,
despite of being born in an extreme environment, has formed with the
slope of a standard Salpeter IMF. The lower mass-limit of the IMF is
not from our models, \revised{but our models (as well as observations)
  show no indication that the IMF should be truncated well above
  $1\msun$ or even $0.5\msun$}. Due to dynamical mass segregation, the
slope of the observed MF is flattened inside a radius of
$0.4\,\pc$. \revised{Outside this radius, our models predict a slope
  even steeper that Salpeter ($\Gamma \gap -2$).} The radius of
$0.4\pc$ was an imposed limit from the observational data used, and we
estimate that a limiting radius of $\sim1\,\pc$ would be required for
the observed MF to match the underlying IMF.

We neglected the Galactic potential and also stellar evolution for the
simulations in this paper. Both processes may have affected the
evolution of the cluster only a little bit due to its young age of
only $\sim 2.5\,\myr$. If this is true our best fitting models suggest
that the Arches cluster was born with an initial virial radius of
$0.77 \pm 0.12\,\pc$ and an initial total mass of about $(4.9 \pm 0.8)
\cdot 10^4\,\msun$ assuming a lower mass limit of $0.5\,\msun$ for the
Salpeter IMF. The uncertainties in the lower mass limit give rise to
additional uncertainties in determining the initial cluster mass. The
King model concentration parameter of the best fitting models is $W_0
= 3$, however, more concentrated models with $W_0 = 5-7$ produced also
reasonable fits so that this parameter is also not very well
constrained.

The missing processes mentioned above will be included in a following
paper to get a more realistic model of the Arches cluster. However
this first step was needed in order to reduce the number of free
parameters for these (computationally more expensive) simulations,
which further constrain the dynamical evolution and tidal effects
acting on the Arches cluster, and thereby the initial conditions of
this nearby starburst, such as the cluster mass, the orbital motion
and the IMF at the birth of the cluster.

{\bf Acknowledgements} We thank the referee for helpful suggestions
that improved this paper. SH and SPZ are grateful for the support from
the NWO Computational Science STARE project \#643.200.503 and NWO
grant \#639.073.803.  AS acknowledges funding from the German Science
Foundation (DFG) Emmy-Noether-Programme under grant \mbox{STO
  496-3/1}.

\bibliographystyle{mn2e}   
\bibliography{harfst}

\end{document}